# Exploratory Analysis of Quality Practices in Open Source Domain

Jie Xu (Corresponding author)

Department of Electrical and Computer Engineering, University of Western Ontario

London, Ontario, N6A 5B9 Canada

Tel: 1-519-858-4397    E-mail: jxu89@uwo.ca

Luiz Fernando Capretz

Department of Electrical and Computer Engineering, University of Western Ontario

London, Ontario, N6A 5B9 Canada

Tel: 1-519-661-2111 ext.85482    E-mail: lcapretz@eng.uwo.ca

Danny Ho

NFA Estimation Inc., Richmond Hill, Ontario, L4C 0A2 Canada

E-mail: danny@nfa-estimation.com

**Abstract**

Software quality assurance has been a heated topic for several decades, but relatively few analyses were performed on open source software (OSS). As OSS has become very popular in our daily life, many researchers have been keen on the quality practices in this area. Although quality management presents distinct patterns compared with those in closed-source software development, some widely used OSS products have been implemented. Therefore, quality assurance of OSS projects has attracted increased research focuses.

In this paper, a survey is conducted to reveal the general quality practices in open source communities. Exploratory analysis has been carried out to disclose those quality related activities. The results are compared with those from closed-source environments and the distinguished features of the quality assurance in OSS projects have been confirmed. Moreover, this study suggests potential directions for OSS developers to follow.

## 1. Introduction

Open source software (OSS) has played a more and more important role since it emerged. When being compared with the development of software in closed-source environment, OSS development presents many distinguishable features in plan, schedule, requirements, testing, user involvement, etc. OSS development teams often comprise of developers in a loosely organized form, and the best motivation for them is user participation and self-commitment. The phases of the OSS development life cycle are not defined so clearly, while the management of OSS projects is not governed by strict rules.

When quality assurance for OSS is concerned, the best practices in closed-source environment are not adopted directly. One obvious reason behind it is that the resources are relatively limited for OSS projects. Although quality assurance seems to be weak according to traditional definitions, some well-known OSS products have been developed and are considered to have high quality. This appealing phenomenon has aroused great interest for research on OSS quality assurance.

This paper provides exploratory analysis on current quality activities in the open source community, based on the data collected by means of a survey.

## 2. Related works

Possible strengths and weaknesses of the OSS development have been discussed by many researchers along with its evolvement. Although there is not a definite agreement on the impacts of the practices in OSS projects, it is widely accepted that some unique features lead to high-quality software.

Mockus, Fielding and Herbsleb (2002) conducted an empirical study to disclose general development patterns in OSS. Two projects, Apache and Mozilla, were examined, and some development activities were claimed to be successful accordingly. Also, differences between the two projects were analyzed since they did not exactly follow the same development style.

Schmidt and Porter (2001) analyzed some major aspects of the open source development and discussed the strength of the development model. They emphasized on the contributions of the large user communities, but only two OSS projects were used for the case study.

Michlmayr, Hunt and Probert (2005) interviewed seven open source developers to identify quality practices and problems in OSS projects. They categorized those quality practices into: infrastructure, processes, and documentation. Also, some potential threats on OSS quality were concluded.

Otte, Moreton and Knoell (2008) conducted a survey and explored popular quality practices in OSS projects. The survey covered practices such as development, testing, defect handling, documentation, infrastructure, and quality assurance.

Zhao and Elbaum (2003) prepared a questionnaire and collected 229 responses for analysis. The authors examined general information, process, testing and user participation of the projects, and explored the results while considering the impacts of product size.

A comprehensive review was done to provide insightful understanding in how to achieve quality in the open source development (Aberdour, 2007). It summarized that four areas were critical: sustainable communities, code modularity, project management, and test project management; nevertheless, many issues still remained unsolved.

**3. Exploratory analysis**

*3.1 Data source*

Many OSS development websites have been designed to provide services for OSS developers and help users find software and share their experiences. Some of these websites also publish historic and status information of the OSS projects that use the services, which attract many researchers and software engineers to do analyses and draw conclusions of OSS development. The project data could be obtained through two methods: data dumps provided by the websites or direct web crawling.

SourceForge.net is the largest OSS development website in the world. As of February 2009, there have been more than 230,000 OSS projects registered to use the development services provided by SourceForge.net and more than 2 million registered users involved in the development activities. Because of its popularity we chose project data from SourceForge.net to conduct our exploratory analysis.

Certain SourceForge.net data has been shared with the University of Notre Dame for research purposes and it consists of more than 100 tables in the data dumps. We mainly collected the download counts of the OSS projects from the repository. A project named FLOSSmole (Collaborative collection and analysis of free/libre/open source project data) has been developed to provide data about OSS projects to the public domain (Howison, Conklin, & Crowston, 2006). Web crawling of the most popular OSS hosts, including SourceForge, Rubyforge, Freshmeat, etc., has been performed mostly on monthly basis to collect data from those websites. We concentrated on OSS projects hosted on SourceForge and extracted related project information of status and ranks from FLOSSmole.

*3.2 Project selection*

Since there are too many OSS projects managed on SourceForge.net, we had to develop criteria to select more suitable ones for our research. Projects that have been developed under sufficient attention could present more complete views of their quality characteristics. We suggested two factors to judge the popularity of the software: download count and rank. As a result, our selection was made on OSS projects that have been downloaded for more than 20,000 times and maintained the rank within the top 10,000 in order to obtain sufficient candidate data points from the more popular projects. For reference, the top 200 popular projects on SourceForge all have been downloaded for more than 2.5 million times.

The software development process of OSS projects on SourceForge is separated into six phases: planning, pre-alpha, alpha, beta, production /stable and mature. We mainly focused on OSS projects with mature status, but also included some projects with production /stable status.

Based on the criteria listed above, we selected 1,571 OSS projects from SourceForge to carry out the survey.

*3.3 Questionnaire design and processing*

To obtain other project information from the project administrators, we decided to design a questionnaire and send it to the administrators of the selected OSS projects. Using the registered names of the project administrators gathered from FLOSSmole, we built an email list for the distribution of the questionnaire. To make it simple enough for them to reply, the questionnaire only comprises of 22 multiple choice questions. These questions take into account the project plan and design, quality requirement, personnel, product complexity, testing and related tools, documentation and so on.

To send the questionnaire to the administrators in an automated way, an excel macro was designed to read the entries and then invoke Outlook Express functions to send out emails to the determined receivers with pre-filled content.

We used direct email contact instead of online survey because the latter is difficult to draw the attentions of the target group. The number of responses was not so satisfactory. Only 278 valid responses were received out of 1,571 sent emails. Most of the sent emails may have been treated as junk emails and were ignored. Only a few of the respondents explicitly expressed no interest in this questionnaire.

*3.4 Analysis of the responses*

We chose part of the 22 questions for analysis, that is, 16 questions, which could reflect the development environment in the OSS domain. Other questions were more about the characteristics of the product itself and thus were utilized in other analyses. The detailed information of the 16 questions could be found in the Appendix A.

3.4.1 Plan and design

Question 1 was about development plan or schedule of the project. In Figure 1, the responses showed that more than half of the projects had no specific plan or schedule and less than 10 percent of them had clear plan or schedule. It confirmed the flexibility of time constraints on OSS projects. Less time pressure made software developers more comfortable to deal with quality issues.

In the OSS domain, developers try to release new versions as soon as possible. New releases tend to fix bugs or enhance features, and make the end users more actively involved. From our questionnaire, we found out more than half of the projects maintained release frequency from 4 months to 6 months (Figure 2). As more users participate in the software development, testing strength is built up with their help. This is the key factor for OSS projects to achieve high quality.

Reliability requirements usually determine how much effort has to be devoted for quality assurance. In Figure 3, the distribution illustrated that less than ten percent of the projects had high reliability requirements, which demonstrated that people might still not be confident enough to develop software with high reliability requirements in the OSS domain.

With regard to software design, we prepared a question for modularity consideration (Question 4). As presented in Figure 4, only ten percent of the projects had no consideration of modularity. With good design of modularity, it helps achieve high quality, since modularity is critical to quite a few quality factors.

3.4.2 Personnel

For personnel factors, the developers of those OSS projects seemed to be more experienced, since many of them had to administer the projects. The responses showed more than half of the developers had more than 5 years related experience in software development (Figure 5).

As for personnel change, the results showed that around 70% of the OSS projects were very stable (Figure 6), with less than ten percent personnel changes. The reason behind it could be that most OSS projects were developed by a team of very small size. The experienced developers and stable team members contribute to smooth development of OSS projects.

3.4.3 Testing

Questions 7 to 12 were designed to explore test related information. The distributions of the respective answers are accordingly displayed in Figure 7, Figure 8, Figure 9, Figure 10, Figure 11, and Figure 12. These projects did not follow distinct patterns throughout all the questions. For test plan (Question 7, Figure 7), 25% of the projects had no test plan, 50% had basic test plans and 25% had very clear test plans, with no outstanding choice. As for test tools (Question 8, Figure 8), around 38 percent of the projects used some tools for testing, but mostly for unit testing. These OSS projects also exhibited casual pattern in test coverage levels of source code (Question 9, Figure 9). It seemed that there was no specific coverage requirement for these projects. One reason might be that very few of the OSS projects were with high reliability requirements. Another reason might be that an essential part of testing was actually performed by users, which made the developers less concerned about the coverage achieved by themselves, and very few of them, i.e. around ten percent, really used coverage tools to examine the coverage percentage they achieved (Question 10, Figure 10). Most of them (around 64%) used the bug tracking system provided by SourceForge to record bug information (Question 11, Figure 11). The system makes it easy for the developers to track bugs and fix them, and encourages users to report bug information. Other projects (around 36%) might have their own bug tracking tools or methods, or even not consider it as a serious issue to manage. For the percentage of development effort located for testing (Question 12, Figure 12), around 64% of the projects put no more than 40% of development effort into testing activities, which once again proved that OSS projects did not pay much attention to improve quality, but relied more on user participation in testing.

3.4.4 Users

For questions about user participation (Questions 13, 14), the responses did not indicate absolute high level involvement of user activities. One possibility of the results might be that the respondents had inconsistent definitions regarding 'users'. Some treated them as potential end users and others could only include those who actually contributed to the projects. The distributions of the responses of these two questions could be found in Figure 13 and Figure 14.

3.4.5 Documentation

For questions about documentation (Questions 15, 16), the responses did not present the same pattern between documentations for developers and users. As for developer documentation, more than half of the projects did not prepare particular instructions for them and only around ten percent of them had detailed guidelines. Since the sizes of those OSS teams are relatively small, the administrators might consider that well-prepared documentation for developers was not critical. When the documentation for users is concerned, around 77% of the projects have prepared major instructions for the users, which also proved that it was essential for the OSS projects to attract more users. The distributions of the responses of these two questions could be found in Figure 15 and Figure 16.

**4. Conclusion and Future Works**

In this study we have collected substantial project information in open source domain and analyzed it to disclose common quality practices among these OSS projects.

Firstly, we found that these OSS projects exhibited loosely organized patterns in planning and design. Most of the projects (around 56%) did not have development plan, but for consideration of modularity design, very few of them (10%) ignored it. Presently, not many projects with high reliability requirements turn to the OSS development model, since the confidence still needs to be built up.

Secondly, the development teams of the OSS projects mostly comprised of experienced developers, and the teams usually were very stable during the development process. Consequently, little training was required for the developers to participate in development. This was further proved by the roughly prepared documentation for developers.

Thirdly, the testing and related issues presented no obvious form, but the intensity of testing activities was much weaker than that of closed-source software. The reason might be that most projects were with low or medium reliability requirements, and many developers depended on the users to assure quality.

Lastly, we could not confirm the claim of high user involvement in OSS projects. There could be a misunderstanding of the definition of 'user', and confusion existed between actual software users and project contributors. However, evidences were conclusive that many methods definitely encouraged the participation of users, such as timely releases, well-prepared documentation for users, etc.

Generally, there are still many hidden issues related to software quality to be discovered in open source domain. The suggestions for future works are as follows:

(1) We plan to define appropriate quality indicators to evaluate overall OSS quality levels. Then further investigation on quality practices can be performed to reveal the real good ones, and thus we can provide more convincing evidences to suggest preferable development patterns;

(2) These quality practices can be used for assessment of OSS quality; specifically, we can use ratings of the practices as predictors for the final quality level, if these practices are solidly confirmed to be closely related to the final quality. Our other direction is to build OSS quality estimation models.

**Acknowledgements**

**References**

Aberdour, M. (2007). Achieving quality in open source software. *Institute of Electrical and Electronics Engineers(IEEE) Software. 24*(1), 58-64.

Howison, J., Conklin, M., & Crowston, K. (2006). FLOSSmole: A collaborative repository for FLOSS research data and analyses. *International Journal of Information Technology and Web Engineering*, *1*(3), 17–26.

Michlmayr, M, Hunt, F., & Probert, D. (2005). Quality practices and problems in free software projects. In *Proceedings of the First International Conference on Open Source Systems*, 24-28.

Mockus, A., Fielding, R. T., & Herbsleb, J. D. (2002). Two case studies of open source software development: Apache and Mozilla. *ACM transactions on software engineering and methodology. 11*(3), 309-346.

Otte, T., Moreton, R., & Knoell, H. D. (2008). Applied quality assurance methods under the open source development model. In *Proceedings of the 32nd Annual IEEE International Computer Software and Applications Conference*. 1247-1252.

Schmidt, D. C., & Porter, A. (2001). Leveraging open-source communities to improve the quality & performance of open-source software. In *Proceedings of the 1st Workshop on Open Source Software Engineering*. Toronto, Canada.

Zhao, L., & Elbaum, S. (2003). Quality assurance under the open source development model. *Journal of Systems and Software, 66*(1), 65-75.

**Appendix A: Questionnaire**

1. Is there a specific plan/schedule for the project?
   - (  )A. No schedule
   - (  )B. Somehow clear schedule
   - (  )C. Very clear schedule

2. How often will the project publish new releases (on average)?
   - (  )A. Not sure
   - (  )B. Every year
   - (  )C. Every six months
   - (  )D. Every quarter
   - (  )E. Every month
   - (  )F. Every week

3. Is there any reliability requirement for the project?
   - (  )A. Low: Slight inconvenience or very small losses when fails;
   - (  )B. Nominal: Moderate losses when fails;
   - (  )C. High: High financial losses, or risk to life when fails.

4. How do you deal with modularity in the project?
   - (  )A. No consideration of modularity
   - (  )B. Redesigned during the development stage
   - (  )C. Prepared at the beginning of the development phase
   - (  )D. Clearly designed during the design stage

5. What is the average related software development experience of the developers? (language, application and platform)
   - (  )A. <1 year
   - (  )B. 1-3 years
   - (  )C. 3-5 years
   - (  )D. >5 years

6. What is the percentage of personnel change during the development?
   - (  )A. <10%
   - (  )B. 10% - 20%

    (　)C. 20% - 30%
    (　)D. 30% - 40%
    (　)E. >40%

7. Do you have test plan for the project?
    (　)A. No test plan
    (　)B. Somehow clear plan (basic requirements)
    (　)C. Very clear test plan (test phases, test cases)

8. Do you use any tool for testing?
    (　)A. No
    (　)B. Yes (Name ________)

9. What percentage of source code is covered during testing?
    (　)A. < 20%
    (　)B. 20% - 40%
    (　)C. 40% - 60%
    (　)D. 60% - 80%
    (　)E. > 80%

10. The previous coverage information is derived from:
    (　)A. Rough estimation
    (　)B. Coverage tool (Name ________)

11. Is the total number of bugs recorded correctly in the Bug Tracking System?(If not, please give a number)
    (　)A. No (Number ________)
    (　)B. Yes

12. What percentage of total development effort is used for testing?
    (　)A. < 20%
    (　)B. 20% - 40%
    (　)C. 40% - 60%
    (　)D. 60% - 80%
    (　)E. > 80%

13. How many users are involved in the project?
    (　)A. < 5
    (　)B. 5 - 10
    (　)C. 10 - 50
    (　)D. 50 - 100
    (　)E. > 100

14. What percentage of defects/bugs do users report?
    (　)A. < 20%
    (　)B. 20% - 40%
    (　)C. 40% - 60%
    (　)D. 60% - 80%
    (　)E. > 80%

15. What documentation is used to help new developers get onboard?
    (　)A. No particular documentation
    (　)B. Major guidelines available
    (　)C. Detailed definition of processes and development guidelines available

16. How is the user documentation prepared?
    (　)A. No particular documentation

( )B. Only draft and incomplete version
( )C. Important parts covered
( )D. Detailed and comprehensive

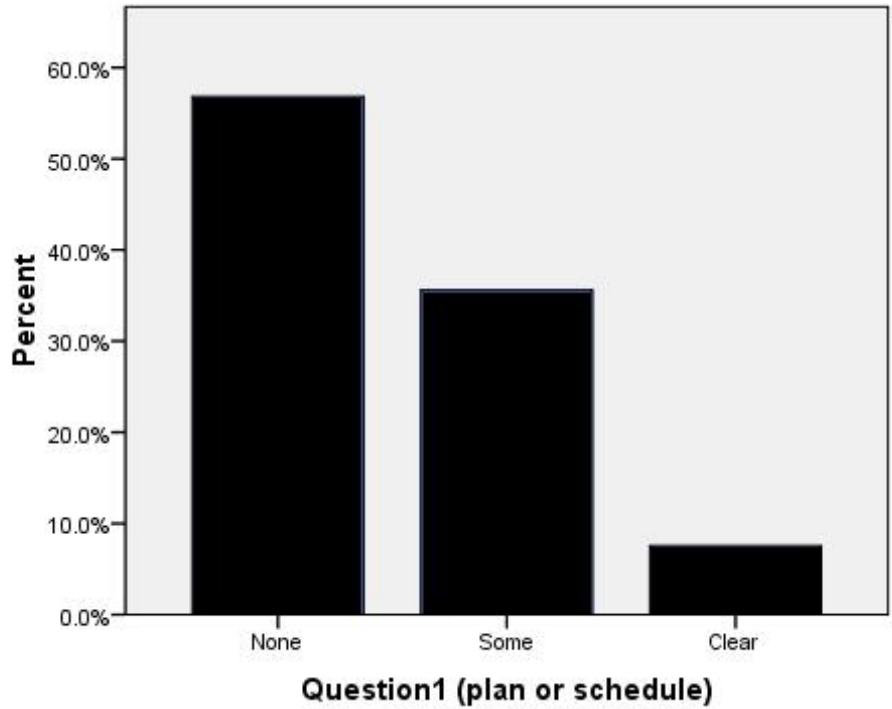

Figure 1. Distribution of Question 1 (plan or schedule)

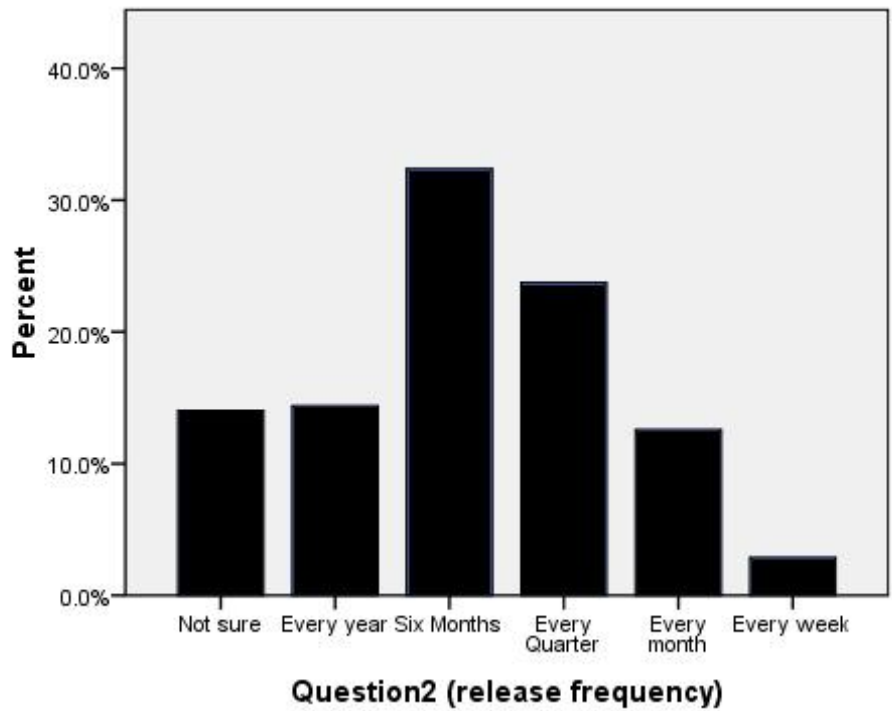

Figure 2. Distribution of Question 2 (release frequency)

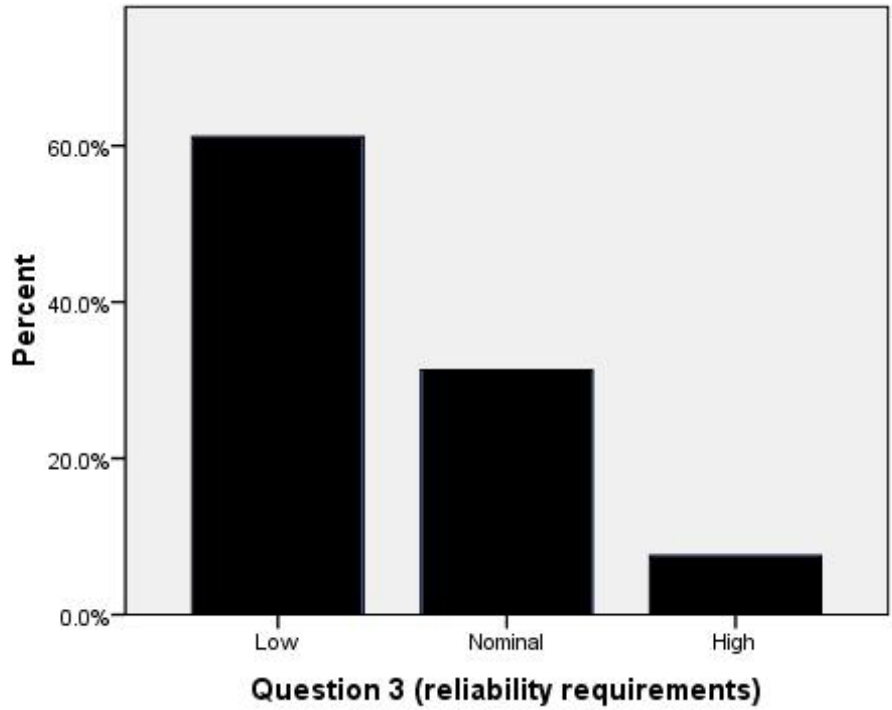

Figure 3. Distribution of Question 3 (reliability requirements)

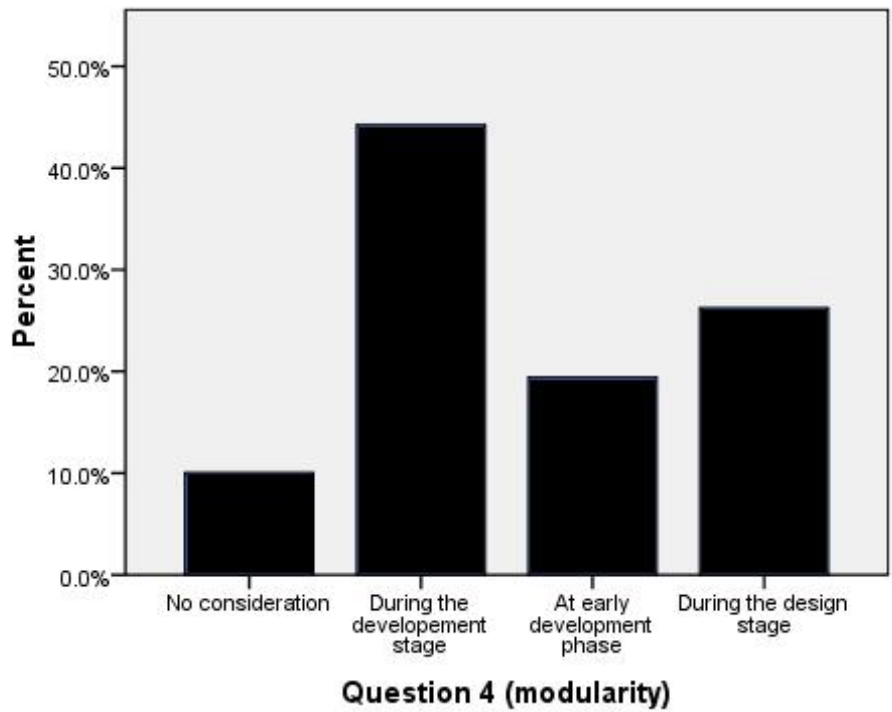

Figure 4. Distribution of Question 4 (modularity)

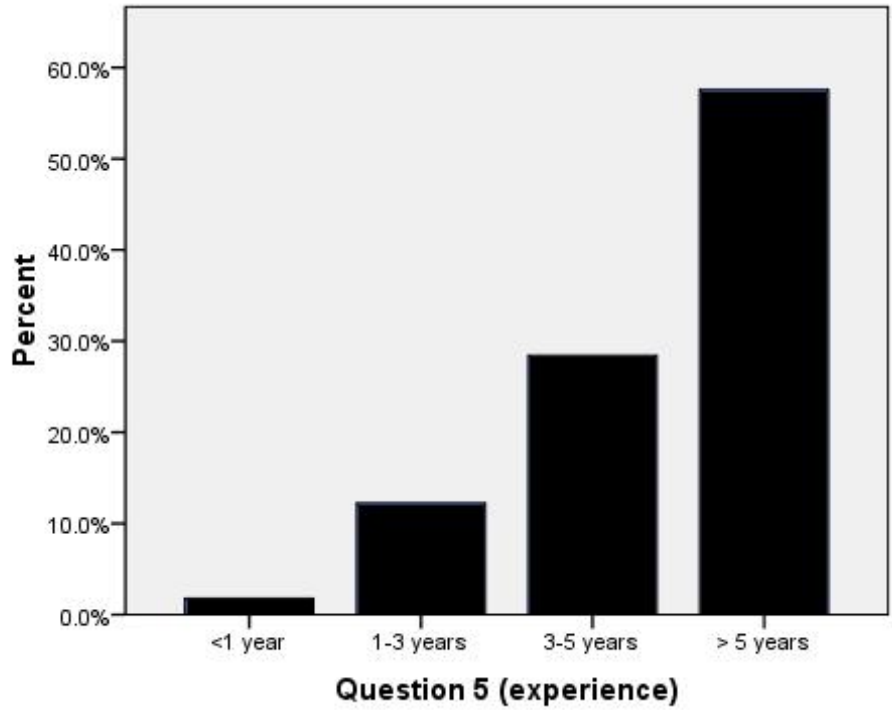

Figure 5. Distribution of Question 5 (experience)

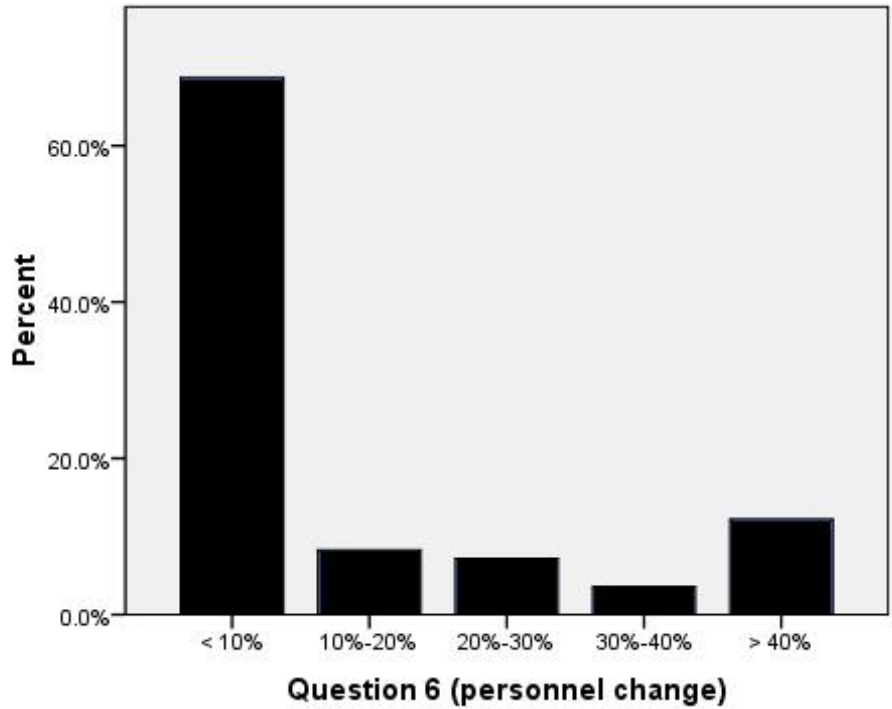

Figure 6. Distribution of Question 6 (personnel change)

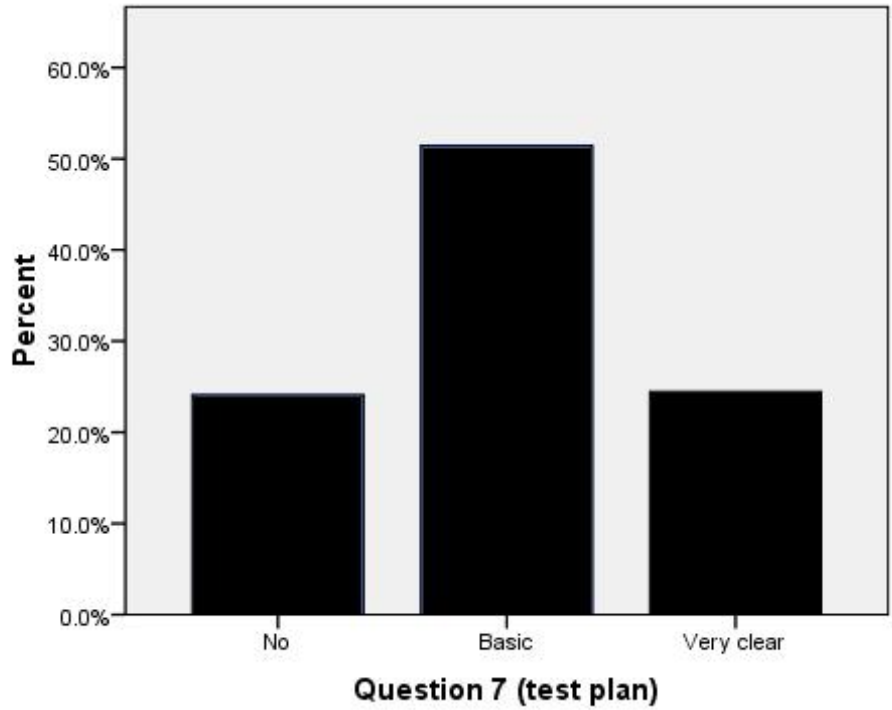

Figure 7. Distribution of Question 7 (test plan)

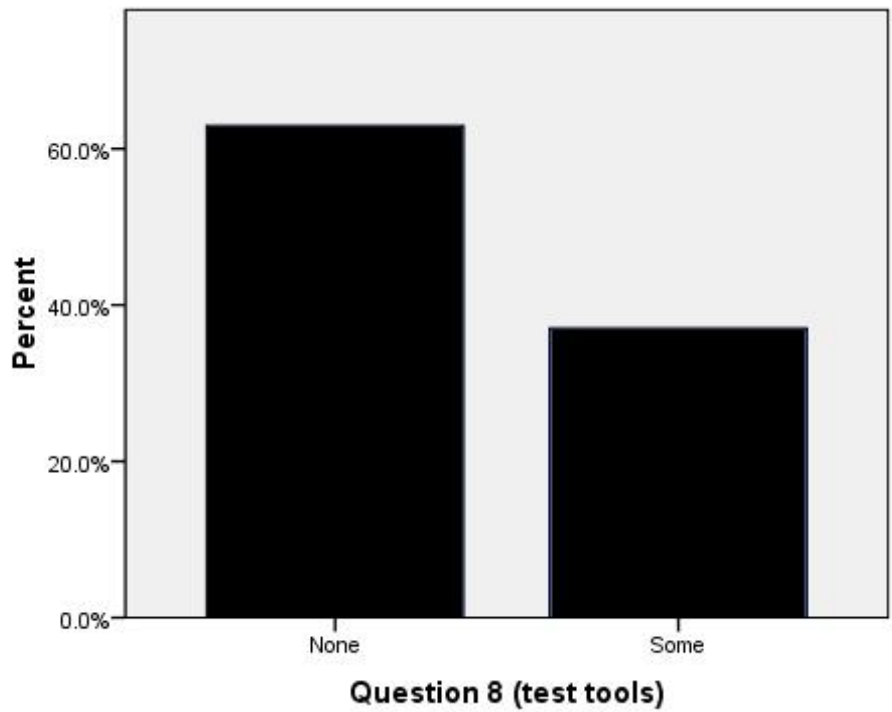

Figure 8. Distribution of Question 8 (test tools)

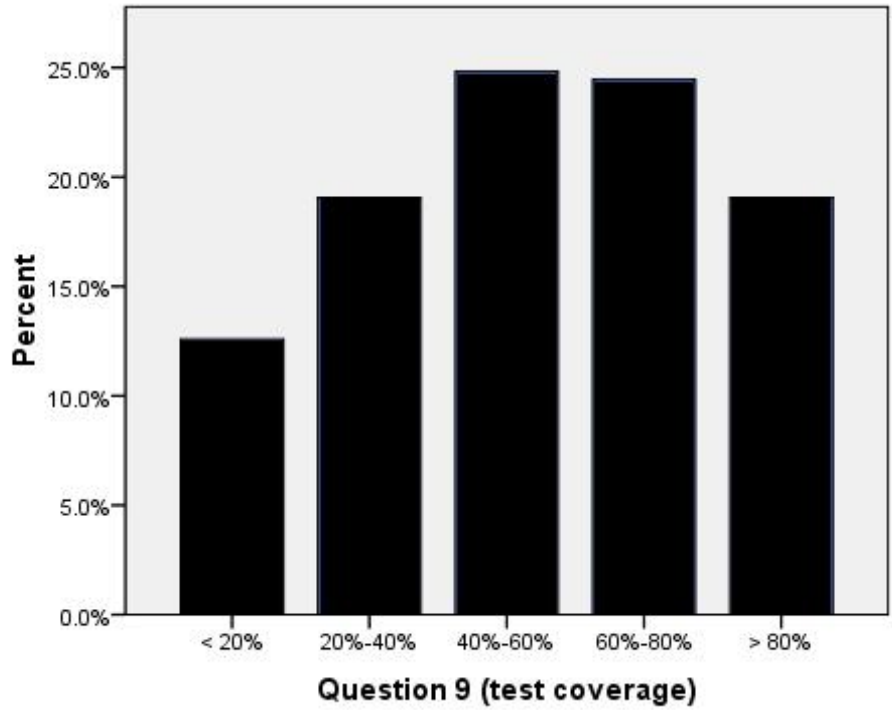

Figure 9. Distribution of Question 9 (test coverage)

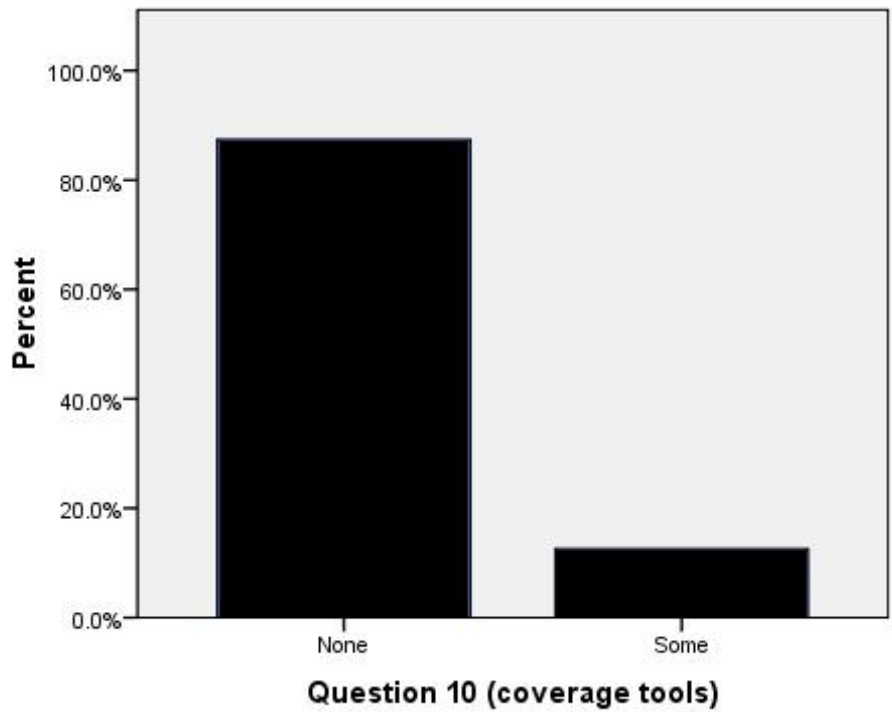

Figure 10. Distribution of Question 10 (coverage tools)

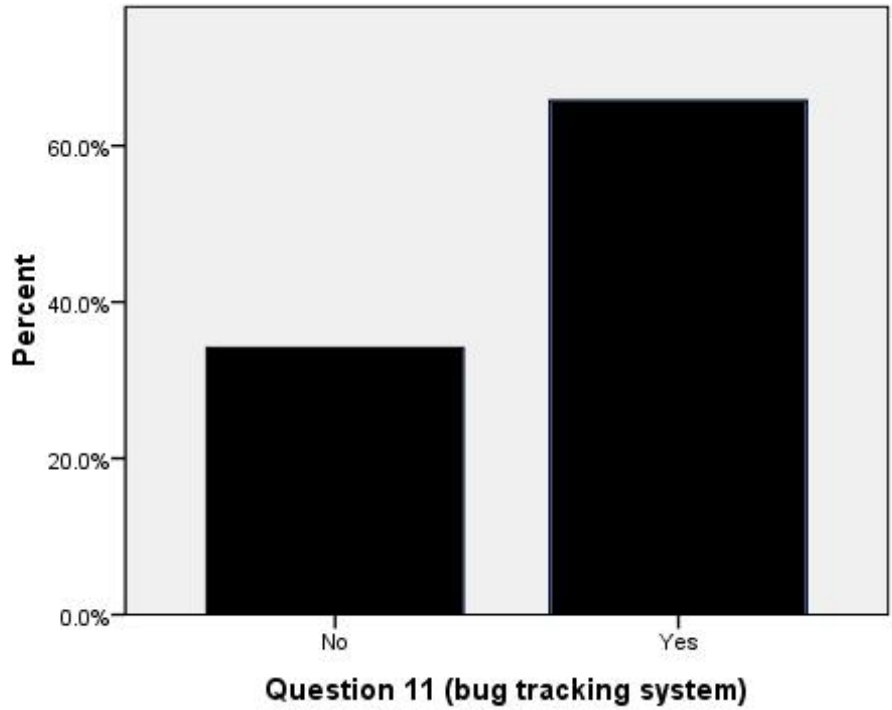

Figure 11. Distribution of Question 11 (bug tracking system)

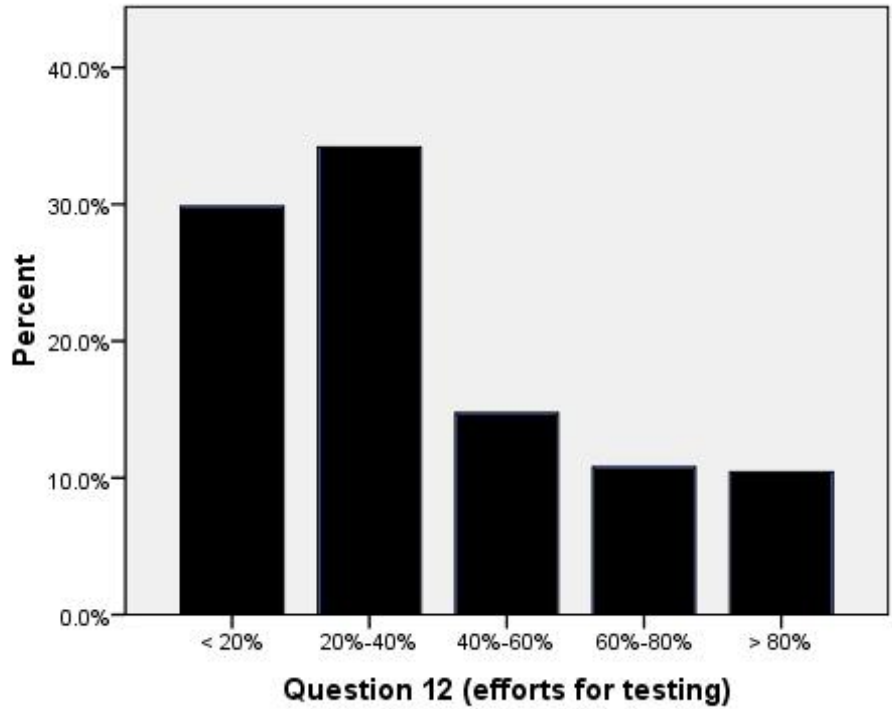

Figure 12. Distribution of Question 12 (effort for testing)

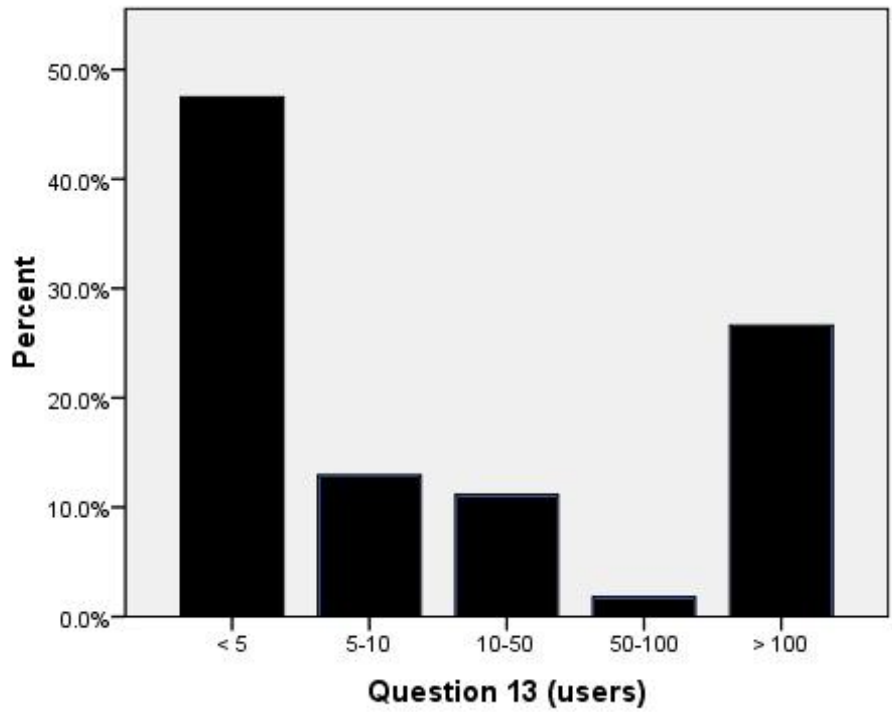

Figure 13. Distribution of Question 13 (users)

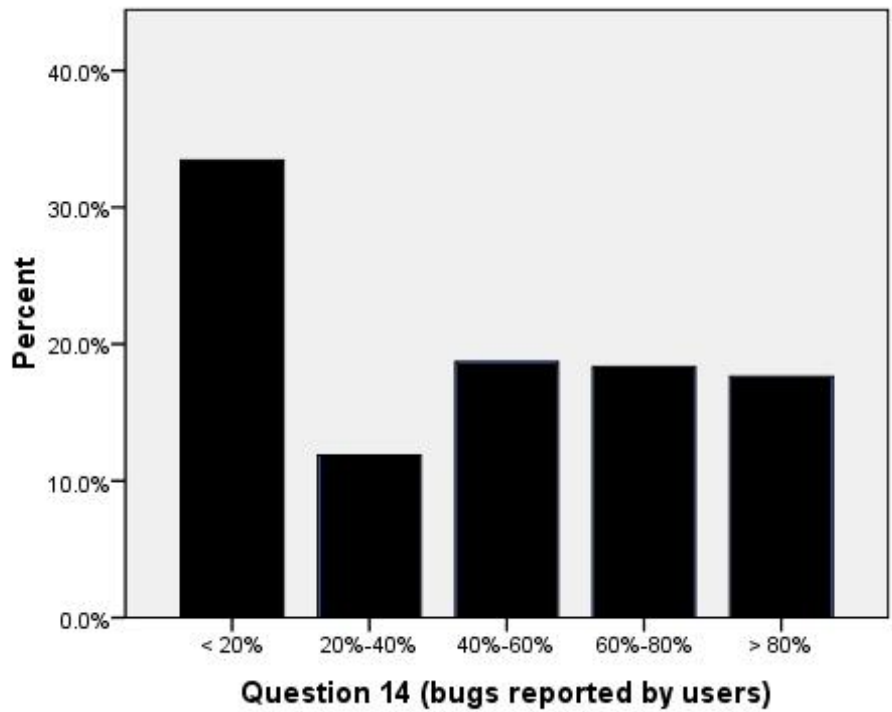

Figure 14. Distribution of Question 14 (bugs reported by users)

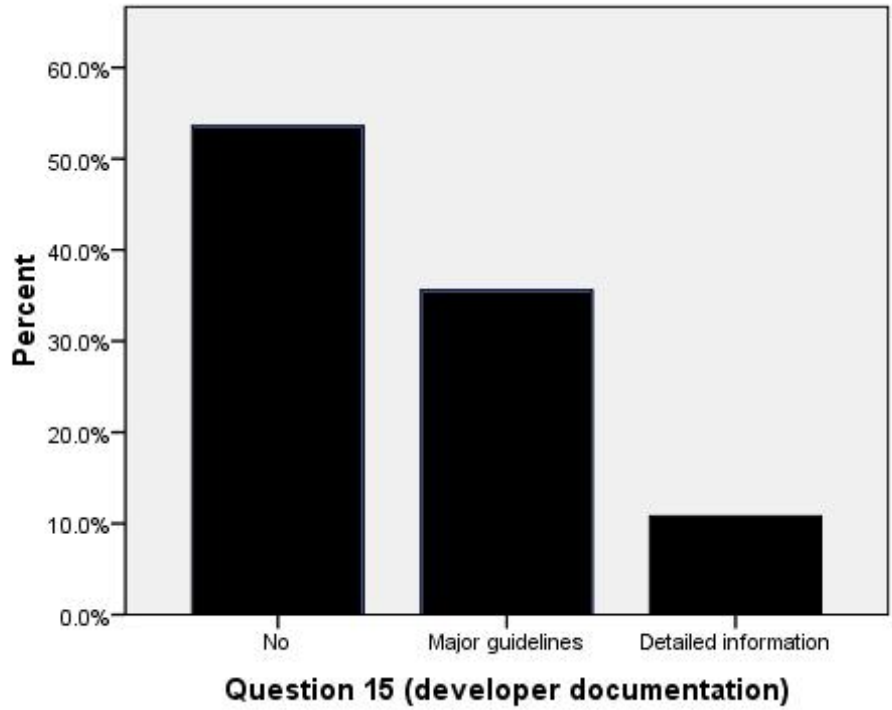

Figure 15. Distribution of Question 15 (developer documentation)

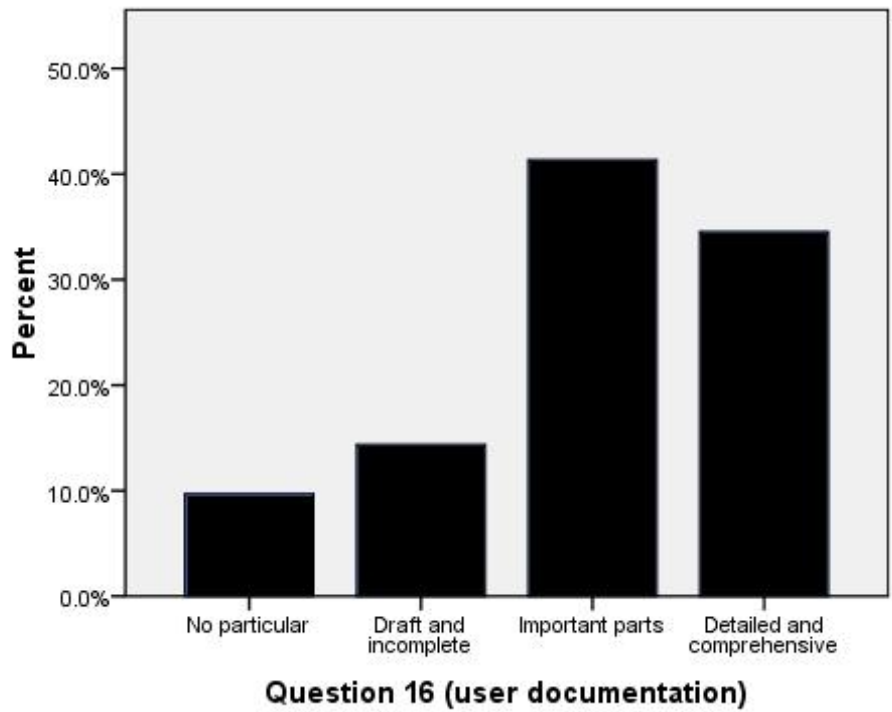

Figure 16. Distribution of Question 16 (user documentation)